\documentclass[runningheads]{llncs}
\usepackage[T1]{fontenc}
\usepackage{graphicx}
\usepackage{booktabs}
\usepackage[misc]{ifsym}
\newcommand{\corr}{(\Letter)}
\usepackage{algorithm}
\usepackage{algorithmicx}
\usepackage{multirow}
\usepackage[table]{xcolor}
\usepackage{amsmath}
\usepackage{amssymb}
\usepackage{pifont}
\usepackage{algpseudocode}
\usepackage{makecell}
\usepackage{array}
\usepackage{hyperref}

\newcommand{\ie}{\emph{i.e.}}

\begin{document}
\title{DiffCold: A Diffusion-based Generative Model for Cold-Start Item Recommendation}
\author{Kangning Zhang\inst{1} \and
Jianghao Lin\inst{1} \corr \and
Yingjie Qin\inst{2} \and
Weinan Zhang\inst{1} \corr \and
Yong~Yu\inst{1} \corr}
\authorrunning{K. Zhang et al.}

\titlerunning{DiffCold: Diffusion-based Generative Model for Cold-Start Recommendation}
%
\institute{Shanghai Jiao Tong University, Shanghai, China\\
\email{\{zhangkangning,linjianghao,wnzhang\}@sjtu.edu.cn}\\
\email{yuyong@apex.sjtu.edu.cn}
\and
Xiaohongshu Inc., Shanghai, China\\
\email{huanling@xiaohongshu.com}}
\maketitle              
\begin{center}
\corr~denotes corresponding authors.
\end{center}

\begin{abstract}
Cold-start item recommendation remains a persistent challenge in real-world systems due to the absence of interaction histories. While prior models attempt to bridge this gap using item content features, they universally suffer from the \textbf{seesaw dilemma}: enhancing performance for cold items inevitably degrades performance for warm items, and vice versa. We identify that this dilemma stems from a fundamental \textbf{distributional disparity}: warm item embeddings occupy a complex ``behavioral manifold" shaped by rich interaction signals, whereas cold item embeddings are constrained to a ``semantic manifold" derived solely from auxiliary content. Existing methods often force a rigid mapping between these inconsistent spaces, causing the model to sacrifice the precision of warm representations to accommodate cold ones.
To address this, we propose \textbf{DiffCold}, a diffusion-based generative model that unifies warm and cold representations. Unlike GANs or VAEs, DiffCold leverages conditional diffusion to reconstruct warm item embeddings from content, preserving the underlying manifold structure without degradation. We further tailor this paradigm with two specific designs: a \textbf{Retrieval-enhanced Aggregator} that initializes generation using semantically similar warm items to bypass inefficient noise, and a \textbf{Simulation-based Representation Alignment} module that enforces distribution consistency between generated and real embeddings via contrastive learning. Experiments on three benchmarks confirm that DiffCold resolves the seesaw dilemma, consistently outperforming state-of-the-art methods across all metrics.
\end{abstract}
\section{Introduction}\label{sec:intro}
In digital platforms~\cite{KZScholarCodeApex,KZScholarProcessRewardSurvey,KZScholarLoopTool,KZScholarSWECycle}, recommender systems~\cite{KZScholarClickPrompt,KZScholarAutoGraphRec} play a crucial role. Embedding-based collaborative filtering models effectively leverage user-item interactions for item recommendation. However, the cold items pose significant challenges due to their absence of historical interaction data, making accurate predictions difficult. To address this, the cold-start recommendation has emerged, combining user-item interaction data with multimedia content to improve representation learning and tackle the cold-start problem.
\begin{figure}[ht]
    \centering
\includegraphics[width=\linewidth]{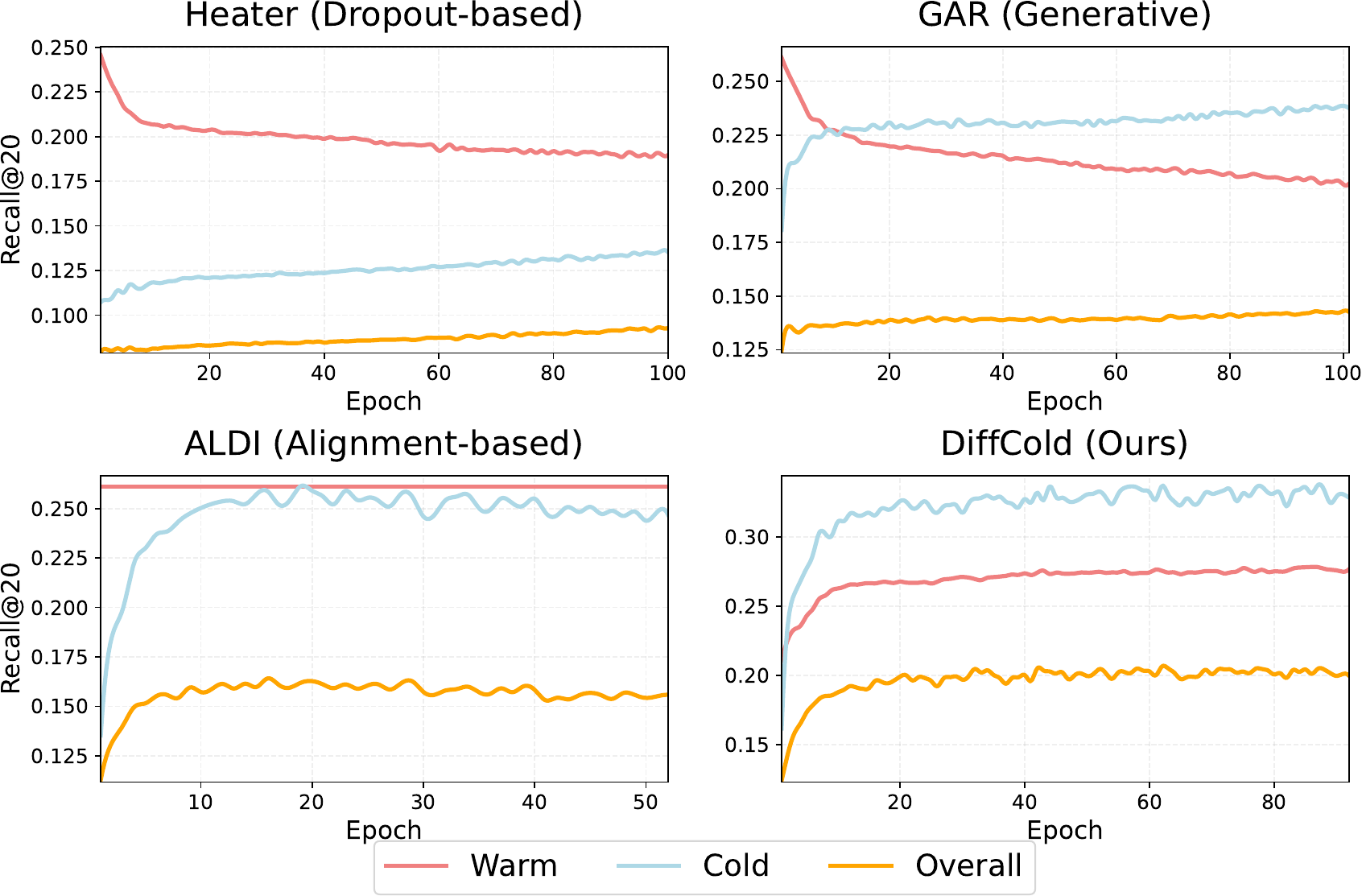}
    \caption{
    The \textbf{seesaw dilemma} for three representative cold-start recommendation methods (\ie, Heater~\cite{Heater}, GAR~\cite{GAR} and ALDI~\cite{ALDI}), where they fail to improve the performance for cold and warm items simultaneously. 
    In contrast, our DiffCold can address the seesaw dilemma and achieve consistent improvement for all three types of task.
    }
    \label{Fig:introduction}
\end{figure}

\begin{table}[ht]
\centering
\caption{The performance comparison relative to the backbone model. Specific experimental data can be referenced in Table \ref{Table:Overall}. ``\ding{51}'' means the performance improvement , while ``\ding{55}'' denotes a decrease. ``---'' indicates no change in performance. More symbols signify a greater rate of change.}
\begin{tabular}{lllll}
\hline
\rowcolor{gray!20}
Groups                           & Methods     & Overall & Cold & Warm \\ \hline
\multirow{2}{*}{Dropout-based}   & DropoutNet&         \ding{55}\ding{55}&      \ding{51}&      \ding{55}\ding{55}\\
                                 & Heater    &         \ding{51}&      \ding{51}&      \ding{55}\\ \hline
\multirow{2}{*}{Generative}  & GAR        &         \ding{51}&      \ding{51}&      \ding{55}\\
                                 & GoRec      &         \ding{55}\ding{55}&      \ding{51}&      \ding{55}\ding{55}\\ \hline
\multirow{2}{*}{Alignment-based} & CLCRec    &         \ding{51}&      \ding{51}&      ---\\
                                 & ALDI      &         \ding{51}&      \ding{51}&      ---\\ \hline
\rowcolor{cyan!15}                        
\textbf{Ours}                             & \textbf{DiffCold}   &         \ding{51}\ding{51}&      \ding{51}\ding{51}&      \ding{51}\\ \hline
\end{tabular}
\label{Table:Introduction}
\end{table}

Despite various attempts—ranging from Dropout-based methods~\cite{DropoutNet,Heater} to Generative~\cite{DeepMusic,GAR,MetaEmb} and Alignment-based approaches~\cite{CLCRec,ALDI}—existing solutions universally suffer from a ``\textbf{seesaw dilemma}``. That is, strategies designed to improve cold-start prediction often come at the cost of degrading warm item performance, and vice versa. We argue that this dilemma arises from a fundamental \textbf{distributional disparity} between the representation spaces of warm and cold items. Warm item embeddings are dynamically learned from interaction signals, forming a complex ``\textbf{behavioral manifold}" that reflects intricate user preferences. In contrast, cold item embeddings are derived solely from content features, forming a distinct ``\textbf{semantic manifold}". Previous methods fail because they attempt to force a rigid alignment between these disparate distributions. For instance, Dropout-based methods force the model to rely on content even for warm items, constraining the flexible behavioral embeddings to a rigid semantic space and hurting accuracy. Similarly, traditional generative models often struggle to capture the multi-modal distribution of the warm manifold, leading to generated cold embeddings that are distributionally inconsistent with the real interaction space. As shown in Figure \ref{Fig:introduction}, we illustrate the recommendation performance of three representative models on the testing set during the training phase. Heater, GAR, and ALDI all struggle to achieve a simultaneous improvement for cold and warm items through the learning process. 

To resolve this dilemma, we propose \textbf{DiffCold}, a diffusion-based generative framework that fundamentally differs from previous mapping or adversarial approaches. We posit that DDPM~\cite{DDPM,DDIM,KZScholarDP3} offers a unique advantage by learning the structure of the warm item manifold through a progressive destruction and reconstruction process. Unlike direct regression or GANs, this mechanism allows DiffCold to decouple the preservation of warm item fidelity from the generation of cold items. Specifically, during the \textit{training phase}, DiffCold perturbs warm item ID embeddings via a forward process that injects variable-scale Gaussian noise. It then learns a reverse denoising process to reconstruct the original behavioral embeddings, using content features as conditional guidance. This establishes a flexible bridge between the semantic and behavioral spaces. In the \textit{inference stage}, DiffCold generates embeddings for cold items by initiating from a starting distribution and progressively refining them through the learned iterative denoising process. However, directly applying this generative paradigm to cold-start recommendation introduces two critical challenges specific to the diffusion process:

(1) \textbf{Starting-point problem: What is the starting representation of cold items at the beginning of inference?} Cold items are unavailable during training, thereby lacking interaction records. Previous works~\cite{DiffRec,DDRM} reveal that using pure Gaussian noise as the starting representation significantly damages the personalized information of items.
Therefore, a crucial question is how to provide a good starting representation during the denoising process for the cold items. In DiffCold, we introduce the \textbf{Retrieval-enhanced Aggregator}. Specifically, for each newly cold-start item, we retrieve the $k$ most similar warm items based on the cosine similarity of item content features and aggregate the representations of these warm items to serve as the starting point during the inference phase. The Retrieval-enhanced Aggregator effectively harnesses the semantic information embedded within content features to facilitate the representation generation for cold items. 

(2) \textbf{Distribution-consistency problem: How can we ensure the consistency of representation distribution between warm and cold items?} 
There is a risk that the generated cold embeddings (during inference) may still diverge from the true distribution of warm embeddings (during training) due to the lack of explicit interaction signals.
Based on this, we introduce a \textbf{Simulation-based Representation Alignment} module.  
In the training phase, we leverage the content features of warm items and the Retrieval-enhanced Aggregator to simulate the generation process of cold items. The resulting simulated representations are explicitly aligned with the representations of warm items through contrastive learning. 
In this way, we ensure the distribution consistency between cold and warm items, thereby solving the seesaw dilemma for cold-start recommendation.
The main contributions of this work are as follows:
\begin{itemize}
    \item We provide a novel perspective on the \textbf{seesaw dilemma}, attributing it to the rigid coupling of semantic and behavioral manifolds, and introduce \textbf{DiffCold}, the first diffusion-based framework specifically tailored to resolve this disparity.
    \item To address the starting-point problem, we design a \textbf{Retrieval-enhanced Aggregator}, which leverages semantic neighbors to provide a robust initialization for the denoising process.
    \item To address the distribution-consistency problem, we propose a \textbf{Simulation-based Representation Alignment} module, which simulates the generation process of cold items using accessible warm items during the training phase. Furthermore, it aligns the representation via contrastive learning, ensuring distributional consistency between cold and warm items.
    \item Comprehensive experiments demonstrate that DiffCold solves the seesaw dilemma and achieves generally SOTA recommendation performance across three datasets. \textbf{Quantitative and visual} analyses further validate the consistency of the representations generated by DiffCold.
\end{itemize}

\section{Related Work}\label{sec:related}

\textbf{Cold-Start Item Recommendation.}
Existing cold-start item recommenders can be categorized into three types: (1) \textit{Dropout-based methods} such as DropoutNet~\cite{DropoutNet}, MTPR~\cite{MTPR}, and Heater~\cite{Heater} randomly drop warm embeddings during training. (2) \textit{Generative methods} utilize content features to generate item ID representations, including GAN-based approaches (GAR~\cite{GAR}), VAE-based methods (GoRec~\cite{GoRec}), and meta-learning techniques (MetaEmb~\cite{MetaEmb}, MWUF~\cite{MWUF}). (3) \textit{Alignment-based methods} minimize the distance between ID embeddings and content features through contrastive learning (CLCRec~\cite{CLCRec}, CCFCRec~\cite{CCFCRec}) or alignment distillation (ALDI~\cite{ALDI,KZScholarAlignRec,KZScholarDREAM,KZScholarDualAligned}).

\noindent\textbf{Diffusion Models for Recommendation.}
Diffusion models, successful in image~\cite{DDPM} and text generation~\cite{DiffuSeq}, are increasingly applied to recommendation systems~\cite{DiffRec_survey}. Applications include data augmentation for sequence recommendation~\cite{DiffuSAR,CaDiRec}, direct sequential modeling~\cite{DreamRec,SeeDRec}, user interaction generation~\cite{DiffRec}, and embedding enhancement~\cite{DDRM}. To our knowledge, DiffCold represents the first application of diffusion models to cold-start item recommendation. By leveraging diffusion's generative capacity to create cold item ID representations aligned with warm item distributions, DiffCold addresses the seesaw dilemma and simultaneously improves recommendation performance for both warm and cold items.

\section{Methodology}

\begin{figure*}[h]
\centering
  \includegraphics[width=\textwidth]{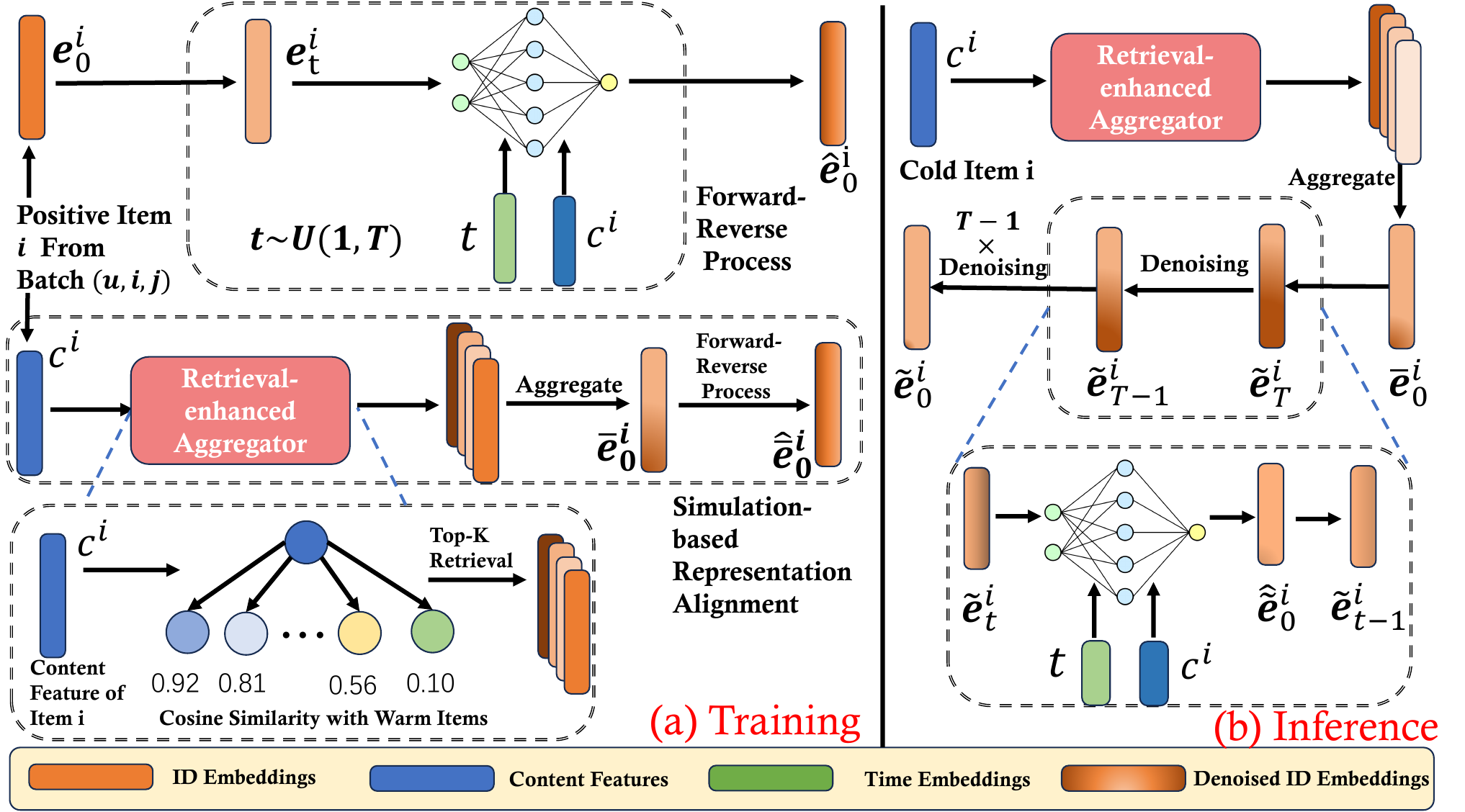}
  \caption{The overall procedure of training and inference in DiffCold. The detailed algorithm of training and inference is summarized in Algorithm \ref{Algorithm:Train} and \ref{Algorithm:Inference}.}
  \label{fig:DiffCold}
\end{figure*}


\subsection{The Framework of DiffCold}
In this section, we elaborate on the framework of DiffCold, as shown in Figure \ref{fig:DiffCold}. The training includes the Forward-Reverse process, the Retrieval-enhanced Aggregator, and the Simulated-based Representation Alignment. During the inference phase, DiffCold employs the Retrieval-enhanced Aggregator to generate the starting representation, followed by a step-by-step denoising process.


\subsection{The Training of DiffCold}\label{sec:train}
Diffusion models mainly consist of forward and reverse processes to model the underlying data distribution. 
For any given embedding-based backbone, we pre-train the embeddings of users and warm items using interaction data. 
During the training process of DiffCold, we adhere to the BPR training paradigm, which involves sampling batches of triplets $(u, i, j) \in \mathcal{B}$, where item $i$ and item $j$ represent the positive item and negative item for the user $u$, respectively. We exclusively apply the forward-reverse process of the diffusion model to the positive item $i$.

\subsubsection{\textbf{Forward-Reverse Process}}\label{sec:Forward-Reverse}
Given the ID embedding of warm item $\textbf{e}^i_0 = \textbf{e}^i$, the forward process gradually adds Gaussian noise to $\textbf{e}_0^i$ with a variance schedule $[\beta_1, \beta_2, \ldots, \beta_T] \in [0,1]$: 
\begin{equation}
\begin{aligned}
    q(\textbf{e}_{1:T}^i | \textbf{e}_0^i) &= \prod_{t=1}^T q(\textbf{e}_t^i | \textbf{e}_{t-1}^i), \\
    \quad q(\textbf{e}_t^i | \textbf{e}_{t-1}^i) &= \mathcal{N}(\textbf{e}_t^i;\, \sqrt{1-\beta_t}\textbf{e}_{t-1}^i,\, \beta_t\mathbf{I}).
\end{aligned}
\end{equation}
where $t$ refers to the time step, $\mathcal{N}$ denotes the Gaussian distribution. If $T \to \infty$, $\textbf{e}_T^i$ approaches a standard Gaussian distribution. 
Following ~\cite{DDRM}, we employing a linear variance noise schedule in the forward process. Let $\alpha_t = 1-\beta_t$, $\bar{\alpha}_t = \prod_{s=1}^t \alpha_s $, $\epsilon \sim \mathcal{N}(\mathbf{0},\mathbf{I})$, we have:
\begin{equation}
\begin{aligned}\label{Eq:forward}
    \textbf{e}_t^i &= \sqrt{\bar{\alpha}}\textbf{e}_0^i + \sqrt{1-\bar{\alpha}_t}\epsilon \\
    1-\bar{\alpha}_t &=s\cdot\left[\alpha_{\min}+\frac{t-1}{T-1}(\alpha_{\max}-\alpha_{\min})\right]
\end{aligned}
\end{equation}
where $s \in (0,1)$ controls the noise scale, $\alpha_{\min}$  and $\alpha_{\max}$ are the minimum and maximum of noise. $t$ is uniformly sampled from $\{1, 2, \ldots, T\}$.

Then we iteratively denoise the noisy item embedding $\textbf{e}^i_t$  in the Reverse process. In order to establish associations between content features and ID embeddings, we employ content features as conditions to guide the reverse process:
\begin{equation}\label{Eq:4}
 p_{\theta}(\textbf{e}_{t-1}^i|\textbf{e}_t^i)=\mathcal{N}(\textbf{e}_{t-1}^i;\mu_\theta(\textbf{e}_t^i,c^i,t),\Sigma_\theta(\textbf{e}_t^i,c^i,t))
\end{equation}
where $c^i$ is the content feature of item $i$. The diffusion model aims to align the distribution $p_{\theta}(\mathbf{e}_{t-1}^i|\mathbf{e}_t^i)$ with the tractable posterior distribution $q(\textbf{e}_{t-1}^i | \textbf{e}_t^i, \textbf{e}_0^i)$, which can be solved in closed form:
\begin{equation}\label{Eq:closed}
    \begin{aligned}
\tilde{\beta}_{t} & =\frac{1-\bar{\alpha}_{t-1}}{1-\bar{\alpha}_t}\beta_t, \\
\tilde{\boldsymbol{\mu}}_t(\mathbf{e}_t^i,\mathbf{e}_0^i) & =\frac{\sqrt{\bar{\alpha}_{t-1}}\beta_t}{1-\bar{\alpha}_t}\mathbf{e}_0^i+\frac{\sqrt{\alpha_t}(1-\bar{\alpha}_{t-1})}{1-\bar{\alpha}_t}\mathbf{e}_t^i, \\
q\left(\mathbf{e}_{t-1}^i\mid\mathbf{e}_t^i,\mathbf{e}_0^i\right) & =\mathcal{N}\left(\mathbf{e}_{t-1}^i;\tilde{\boldsymbol{\mu}}_t\left(\mathbf{e}_t^i,\mathbf{e}_0^i\right),\tilde{\beta}_t\mathbf{I}\right), \\
\end{aligned}
\end{equation}
Therefore, $\Sigma_\theta({\mathbf{e}}_t^i,c^i,t)$ in Eq. (\ref{Eq:4}) is set as $\tilde{\beta}_{t}  =\frac{1-\bar{\alpha}_{t-1}}{1-\bar{\alpha}_t}\beta_t$, $\mu_\theta({\mathbf{e}}_t^i,c^i,t)$  can be factorized as:
\begin{equation}
\begin{aligned}\label{Eq:one-prediction}
\mu_\theta({\mathbf{e}}_t^i,c^i, t)&=\frac{\sqrt{\bar{\alpha}_{t-1}}\beta_t}{1-\bar{\alpha}_t}\hat{\mathbf{e}}_0^i+\frac{\sqrt{\alpha_t}(1-\bar{\alpha}_{t-1})}{1-\bar{\alpha}_t}{\mathbf{e}}_t^i, \\
    \hat{\mathbf{e}}_0^i &=  f_{\theta} (\mathbf{e}_t^i, c^i, t)
\end{aligned}
\end{equation}
where $\hat{\mathbf{e}}_0^i$ denotes the model's prediction for the original ID embedding based on the noisy embeddings $\mathbf{e}^i_t$, content feature $c^i$ and the time step $t$. We apply a reparameterization technique that allows the diffusion model to directly output $\hat{\mathbf{e}}_0^i$ as an estimate of the original item embedding $\mathbf{e}^i_0$. This simplifies the objective to a basic reconstruction loss term:
\begin{equation} \label{Eq:recon}
    L_{\text{recon}} = \mathbb{E}_{\mathbf{e}^i_0, t\sim\mathcal{U}(1,T)} \left[||\mathbf{e}_0^i- \hat{\mathbf{e}}^i_0||^2\right]
\end{equation}

Furthermore, the predicted item embedding $\hat{\mathbf{e}}_0^i$ should also satisfy the requirements of the recommendation task. Therefore, we synchronously optimize it using the BPR loss:
\begin{equation}\label{Eq:bpr1}
    L_{\text{bpr-cold}} = \sum_{(u,i,j) \in \mathcal{B}} -\log(\sigma(\mathbf{e}^u\cdot \hat{\mathbf{e}}^i_0- \mathbf{e}^u\cdot \mathbf{e}^j))
\end{equation}

\subsubsection{\textbf{Retrieval-enhanced Aggregator}}\label{sec:Similarity-aware}
Previous works usually overlook the deeper semantic associations contained within the content features when generating the cold ID embeddings. In DiffCold, we introduce the Retrieval-enhanced Aggregator, which leverages the cosine similarity between item to model semantic associations~\cite{KZScholarTokenCrossing} and retrieve relevant items.

Specifically, we pre-calculate the similarity score $s^{i,j}$ for the item pair $(i,j) \in \mathcal{I}$ by measuring the cosine similarity between their original content features:
\begin{equation}
    s^{i,j} = \frac{(c^i)^Tc^j}{\|c^i\|\|c^j\|}
\end{equation}
For each item $i \in \mathcal{I}$, we maintain a set of the $k$ most similar \textbf{warm items}, forming the retrieval set. The aggregated semantic embeddings could be obtained through mean aggregation:
\begin{equation}\label{Eq:retrieval_set}
\begin{aligned}
    \mathcal{I}^i_R &= \{j \in \mathcal{I}_w| s^{i,j} \in \text{top-}K(s^{i,\cdot})\}, \\
    \bar{\mathbf{e}}^i_0 &= \frac{1}{|\mathcal{I}^i_R|} \sum_{j \in \mathcal{I}_R^i} \mathbf{e}^j
\end{aligned}
\end{equation}


\subsubsection{\textbf{Simulated-based Representation Alignment}}\label{sec:Simulation-based}
The core of DiffCold lies in generating cold embeddings that align with the distribution of warm items through the content features. However, during the training phase, cold items are inaccessible, which contributes to the distribution inconsistency between warm and cold items observed in previous work and the see-saw dilemma. In response to this, we propose the \textbf{Simulated-based Representation Alignment}. Specifically, for warm item $i$, we utilize the Retrieval-enhanced Aggregator to generate the simulated starting point $\bar{\textbf{e}}^i_0$ \eqref{Eq:retrieval_set} and the Forward-Reverse process for simulated representation:
\begin{equation}\label{Eq:simulation}
\begin{aligned}
      \bar{\mathbf{e}}^i_t = \sqrt{\bar{\alpha}_t}\bar{\mathbf{e}}_0^i+\sqrt{1-\bar{\alpha}_t}\boldsymbol{\epsilon}, \quad
      \hat{\bar{\mathbf{e}}}^i_0 = f_\theta(\bar{\mathbf{e}}^i_t,c^i,t),
\end{aligned}
\end{equation}



Subsequently, we align the simulated representations  $ \hat{\bar{\mathbf{e}}}^i_0$ with the real warm item representations $\textbf{e}^i$, by leveraging InfoNCE loss to maximize the mutual information:
\begin{equation}\label{Eq:simulation-align}
   L_{\text{sim-align}} = \sum_{i \in \mathcal{B}} -\text{log}\frac{\exp(\hat{\bar{\mathbf{e}}}^i_0 \cdot \mathbf{e}^i / \tau)}{\sum_{j \in \mathcal{B}} \exp( \hat{\bar{\mathbf{e}}}^i_0 \cdot \mathbf{e}^j / \tau)}
\end{equation}


\begin{algorithm}
\caption{The Training of DiffCold}\label{Algorithm:Train}
\begin{algorithmic}[1]  
\Require Interaction matrix $R$, pre-trained user and warm item embedding table $E^{\mathcal{U}}$ and $E^{\mathcal{I}_w}$, diffusion step $T$, parameters of diffusion model $\theta$. 
\Repeat
    \State Randomly sample a batch of triplets $(u, i, j)$. 
    \ForAll {$\mathbf{(u, i, j)} \in \mathcal{B}$}
        \State Sample $t \sim \mathcal{U}(1, T)$;
        \State Compute $\mathbf{e}^i_t$ given $\mathbf{e}_0^i$ and $t$ via Eq.~\eqref{Eq:forward}; 
        \State Calculate the predicted $\hat{\mathbf{e}}^i_0$ via Eq.~\eqref{Eq:one-prediction};
        \State Calculate the predicted $\hat{\bar{\textbf{e}}}_0^i$ via Eq.~\eqref{Eq:simulation};
        \State Calculate $\mathcal{L}_{\mathrm{overall}}$ by Eq.~\eqref{Eq:overall}; 
        \State Update $\theta$, $E^\mathcal{U}$, $E^{\mathcal{I}_w}$ via gradient descent;
    \EndFor
\Until{converged}
\Ensure Optimized $\theta$, $E^\mathcal{U}$, $E^{\mathcal{I}_w}$.
\end{algorithmic}
\end{algorithm}

\subsubsection{\textbf{Warm-augment Representation Learning}}\label{sec:warm-aug}
The optimization for cold items could cause a shift in the distribution of the original warm items and user embeddings, which adversely affects the recommendation performance for warm items. Based on this observation, we further align the representation between users and preferred warm items. Specifically, $L_{\text{bpr2}}$ optimizes personalized ranking and $L_{warm-aug}$ aligns global distributions via contrastive learning:
\begin{equation}
    \begin{aligned}
        L_{\text{bpr-warm}} &=  \sum_{(u,i,j) \in \mathcal{B}} -\log(\sigma(\mathbf{e}^u\cdot \mathbf{e}^i- \mathbf{e}^u\cdot \mathbf{e}^j)), \\
        L_{warm-aug}  &=   \sum_{i \in \mathcal{B}} -\text{log}\frac{\exp(\mathbf{e}^u \cdot \mathbf{e}^i / \tau)}{\sum_{j \in \mathcal{B}} \exp( \mathbf{e}^u \cdot \mathbf{e}^j / \tau)},
    \end{aligned}
\end{equation}

\subsubsection{\textbf{Multi-task learning of Representations for Warm and Cold Items}}
Based on the aforementioned procedures, we categorize the training constraints of DiffCold into three classes: the reconstruction loss for diffusion model training, the loss terms serving the denoised representation, and the loss term for warm representations, respectively:
\begin{equation}\label{Eq:overall}
\begin{aligned}
    L_{\text{overall}}  &= L_{recon} + \underbrace{\alpha_1L_{\text{bpr-cold}}  +  \beta_1L_{\text{sim-align}}}_{\text{Denoised Representations}} \\
     &+ \underbrace{\alpha_2L_{\text{bpr-warm}} + \beta_2L_{\text{warm-aug}}}_{\text{Warm Item Representations}},
\end{aligned}
\end{equation}
where $L_{recon}$ is calculated according to Eq \eqref{Eq:recon}, $\alpha_1$, $\alpha_2$ and $\beta_1$, $\beta_2$ are hyperparameters that balance the weight of different loss terms.

\subsection{The Inference of DiffCold} \label{sec:inference}

\begin{algorithm}
\caption{The Inference of DiffCold}\label{Algorithm:Inference} 
\begin{algorithmic}[1] %
\Require Optimized user and warm item embeddings $E^\mathcal{U}$, $E^{\mathcal{I}_w}$, optimized diffusion model $\theta$, diffusion step $T$.
    \ForAll {$i \in \mathcal{I}_c$}
        \State Calculate the starting representation $\bar{\mathbf{e}}_0^i$ via Eq.~\eqref{Eq:retrieval_set};
        \State Calculate $\tilde{\mathbf{e}}_0^i$ via Eq.~\eqref{Eq:inference};
    \EndFor
    \ForAll {$u \in \mathcal{U}$}
        \State Calculate preference score for warm and cold items: $\hat{y}_{ui}^{(w)} = (e^u)^\top e^i, \forall i \in \mathcal{I}_w \setminus \mathcal{I}_u$ and $\hat{y}_{ui}^{(w)} = (e^u)^\top \tilde{\mathbf{e}}^i_0, \forall i \in \mathcal{I}_c$
    \EndFor
\Ensure $\{\hat{y}_{ui}^{(w)}\} \cup \{\hat{y}_{ui}^{(c)}\}$ across all users.
\end{algorithmic}
\end{algorithm}

During the inference stage, we focus on employing DiffCold to generate cold item embeddings. Diffusion typically uses Gaussian noise as the starting point for denoising. However, pure Gaussian noise significantly impacts the personalization information of items~\cite{DiffRec,DDRM,KZScholarFints}. Therefore, we first utilize the Retrieval-enhanced Aggregator to generate the inference starting representation $\bar{\textbf{e}}^i_0$ for each cold item $i \in \mathcal{I}_c$ as shown in Eq \eqref{Eq:retrieval_set}. Then, the T-step noisy embeddings $\bar{\mathbf{e}}^i_T$ can be calculated:
\begin{equation}
    \bar{\mathbf{e}}^i_T =\sqrt{\bar{\alpha}_T}\bar{\mathbf{e}}^i_0+\sqrt{1-\bar{\alpha}_T}\boldsymbol{\epsilon}
\end{equation}
Subsequently, we iteratively execute the denoising process to generate the cold-start item embedding conditioned on the step embedding and content feature $c^i$:
\begin{equation}\label{Eq:inference}
\begin{aligned}
\hat{\tilde{\mathbf{e}}}_0^i &= f_{\theta} (\tilde{\mathbf{e}}_t^i, c^i, t) \\
\tilde{\mathbf{e}}_{t-1}^i &=\frac{\sqrt{\alpha_t}(1-\bar{\alpha}_{t-1})}{1-\bar{\alpha}_t}\tilde{\mathbf{e}}_t^i+\frac{\sqrt{\bar{\alpha}_{t-1}}\beta_t}{1-\bar{\alpha}_t}\hat{\tilde{\mathbf{e}}}_0^i,
\end{aligned}
\end{equation}
where $t$ executes in descending order from $T$ to 1, and $\tilde{\mathbf{e}}^i_T = \bar{\mathbf{e}}_T^i$. The output $\tilde{\mathbf{e}}_{0}^i$ is utilized as the generated ID embeddings of cold items.

\section{Experiments}

\begin{table}[ht]
\caption{Statistics of the experimental datasets.}\label{Table:Statistics of datasets}
\centering
\begin{tabular}{c|c|c|c|c|c|c}
\hline
\rowcolor{gray!20} Datasets & \# Users & \makecell{\# Warm \\ Items} &  \makecell{\# Cold \\ Items} & \makecell{\# Inter \\ -actions} & \makecell{\# Content \\ Feature} & Sparsity \\ \hline
Movielens & 6040 & 2964 & 742 & 1000209 & 206 & 95.53\% \\
Citeulike & 5551 & 13584 & 3396 & 204986 & 300 & 99.78\% \\
Xing & 106881 & 16415 & 4104 & 3856580 & 2738 & 99.82\% \\ \hline
\end{tabular}
\end{table}
\begin{table}[ht]
\centering
\caption{Overall, Cold and Warm recommendation performance of all models. The R@20 and N@20 mean Recall@20 and NDCG@20, respectively. The improvements are calculated by comparing DiffCold to the best baseline (\underline{underline}) on each dataset.
}
    \resizebox{\linewidth}{!}{
\begin{tabular}{cc|cccccc|cccccc|cccccc}
\hline
\rowcolor{gray!20}
\multicolumn{2}{c|}{}                                       & \multicolumn{6}{c|}{\textbf{Movielens}}                                                                                                                                                                 & \multicolumn{6}{c|}{\textbf{Citeulike}}                                                                                                                                                                 & \multicolumn{6}{c}{\textbf{Xing}}                                                                                                                                                                    \\
\cline{3-20}
\rowcolor{gray!20}
\multicolumn{2}{c|}{}                                       & \multicolumn{2}{c|}{\textbf{Overall}}                              & \multicolumn{2}{c|}{\textbf{Cold}}                                 & \multicolumn{2}{c|}{\textbf{Warm}}                            & \multicolumn{2}{c|}{\textbf{Overall}}                              & \multicolumn{2}{c|}{\textbf{Cold}}                                 & \multicolumn{2}{c|}{\textbf{Warm}}                            & \multicolumn{2}{c|}{\textbf{Overall}}                              & \multicolumn{2}{c|}{\textbf{Cold}}                                 & \multicolumn{2}{c}{\textbf{Warm}}                             \\
\rowcolor{gray!20}
\multicolumn{2}{c|}{\multirow{-3}{*}{\textbf{Methods}}}      & \textbf{R@20}               & \multicolumn{1}{c|}{\textbf{N@20}} & \textbf{R@20}               & \multicolumn{1}{c|}{\textbf{N@20}} & \textbf{R@20}               & \textbf{N@20}                 & \textbf{R@20}               & \multicolumn{1}{c|}{\textbf{N@20}} & \textbf{R@20}               & \multicolumn{1}{c|}{\textbf{N@20}} & \textbf{R@20}               & \textbf{N@20}                 & \textbf{R@20}               & \multicolumn{1}{c|}{\textbf{N@20}} & \textbf{R@20}               & \multicolumn{1}{c|}{\textbf{N@20}} & \textbf{R@20}               & \textbf{N@20}                 \\ \hline\hline
\rowcolor{yellow!10}                 &  \textbf{Backbone}    & { 0.1044} & { 0.1457}      & { 0.0290} & { 0.0300}      & { 0.2309} & { 0.1924} & { 0.1143} & { 0.0948}      & { 0.0047} & { 0.0022}      & { 0.2509} & { 0.1502} & { 0.1920} & { 0.1684}      & { 0.0057} & { 0.0028}      & { 0.4566} & { 0.2907} \\ \cline{2-20}
                                     & \textbf{DropoutNet}  & { 0.0673} & { 0.1022}      & { 0.0448} & { 0.0567}      & { 0.1474} & { 0.1318} & 0.0392                        & 0.0293                             & 0.0768                        & 0.0435                             & 0.0829                        & 0.0460                        & 0.1273                        & 0.1021                             & 0.2912                        & 0.2051                             & 0.2298                        & 0.1432                        \\
                                     & \textbf{Heater}      & 0.1074                        & 0.1536                             & 0.1270                        & 0.1151                             & 0.1678                        & 0.1549                        & { 0.1117} & { 0.0925}      & { 0.1710} & { 0.0936}      & { 0.2450} & { 0.1466} & 0.2000                        & 0.1714                             & 0.1227                        & 0.1103                             & 0.4449                        & 0.2830    \\
                                     & \textbf{MetaEmb} &  0.1044 & 0.1371 & 0.1457 & 0.1483 & 0.2309 & 0.1924 &   0.1188 & 0.0960 & 0.1904 & 0.1076 & 0.2509 & 0.1502 & 0.1920 & 0.1684 & 0.3227 & 0.2066 & 0.4566 & 0.2907 \\
                                     & \textbf{GAR}         & \underline {0.1270}                  & \underline {0.1881}                       & 0.1924                        & 0.1955                             & 0.2269                        & 0.2044                        & { 0.1426} & { 0.1086}      & { 0.2421} & { 0.1429}      & { 0.1981} & { 0.1185} & 0.2432                        & 0.1845                             & 0.3075                        & 0.2004                             & 0.4088                        & 0.2519                        \\
                                     & \textbf{GoRec}       & 0.0713  & 0.1087        & 0.1921    & \underline{0.1966}   & 0.0994    & 0.0841   & 0.0680                        & 0.0511                             & 0.2127                        & 0.1222                             & 0.0896                        & 0.0482                        & 0.1360                        & 0.1035                             & 0.2226                        & 0.1349                             & 0.2450                        & 0.1461                        \\
                                     & \textbf{CLCRec}      & 0.1053                        & 0.1466                             & 0.1499                        & 0.1506                             & 0.2309                        & 0.1924                        & 0.1596                  & 0.1201                      & 0.2674                 & \underline {0.1545}                       & 0.2509                        & 0.1502                        & \underline {0.2622}                  & \underline {0.2239}                       & \underline {0.3293}                  & \underline {0.2284}                       & 0.4566                        & 0.2907                        \\
                                     & \textbf{ALDI}        & 0.1203                        & 0.1671                             & \underline {0.1959}                  & 0.1938                             & 0.2309                        & 0.1924                        & 0.1495                        & 0.1118                             & 0.2311                        & 0.1286                             & 0.2509                        & 0.1502                        & 0.2583                        & 0.2089                             & 0.2900                        & 0.1985                             & \underline{0.4566}                       & \underline{0.2907}                        \\
                                     & \textbf{PAD-CLRec} &  0.1260 & 0.1704 & 0.1874 & 0.1832 & \underline{0.2386} & \underline{0.2027} & \underline{0.1676} & \underline{0.1345} & \underline{0.2861} & 0.1537 & \underline{0.2611} & 0.1579 & 0.2519 & 0.2043 & 0.2975 & 0.2059 & 0.4553 &0.2891\\
                           \cline{2-20}
\rowcolor{cyan!15}                   &  \textbf{DiffCold (Ours)}    & \textbf{ 0.1463}   & \textbf{0.1966}       & \textbf{0.2148}   & \textbf{0.2052}         & \textbf{0.2593}   & \textbf{0.2225}  & \textbf{0.2054}                        & \textbf{0.1593 }                            & \textbf{0.3295}                        & \textbf{0.1953}                             & \textbf{0.2873}                        & \textbf{0.1797}                      & \textbf{0.2844}                        & \textbf{0.2309}                             & \textbf{0.3657}                        & \textbf{0.2594}                            & \textbf{0.4775}                        & \textbf{0.3064}                       \\
\rowcolor{cyan!15}\multirow{-11}{*}{\textbf{MF}}       &  \textbf{Improv. (\%)} & 15.20\%                        & 4.52\%                              & 9.65\%                         & 4.37\%                              & 8.68\%                        & 9.77\%                        & 22.55\%                        & 18.44\%                             & 15.17\%                        & 26.41\%                             & 10.03\%                         & 13.80\%                         & 8.47\%                         & 3.13\%                              & 11.05\%                        & 13.57\%                             & 4.58\%                         & 5.40\%                         \\ \hline\hline
\rowcolor{yellow!10}                 &  \textbf{Backbone}    & { 0.1269} & { 0.1871}      & { 0.0266} & { 0.0254}      & { 0.2805} & { 0.2465} & { 0.1190} & { 0.0999}      & { 0.0083} & { 0.0044}      & { 0.2611} & { 0.1591} & { 0.2033} & { 0.1803}      & { 0.0123} & { 0.0075}      & { 0.4832} & { 0.3110} \\ \cline{2-20}

                                     & \textbf{DropoutNet}  & { 0.0871} & { 0.1234}      & { 0.0450} & { 0.0511}      & { 0.1905} & { 0.1639} & { 0.0527} & { 0.0384}      & { 0.1135} & { 0.0645}      & { 0.1135} & { 0.0611} & { 0.1337} & { 0.1119}      & { 0.2158} & { 0.1519}      & { 0.2885} & { 0.1809} \\
                                     & \textbf{Heater}      & 0.1262                        & 0.1850                             & 0.0498                        & 0.0489                             & 0.2783                        & 0.2436                        & 0.1177                        & 0.0987                             & 0.1333                        & 0.0776                             & 0.2589                        & 0.1568                        & { 0.2126} & { 0.1842}      & { 0.1331} & { 0.1177}      & { 0.4743} & { 0.3049} \\
                                     & \textbf{MetaEmb} & 0.1269 & 0.1871 & 0.1523 & 0.1536 & 0.2805 & 0.2465 & 0.1219 & 0.1008 & 0.2116 & 0.1185 & 0.2611 & 0.1591 & 0.2204 & 0.1885 & 0.3102 & 0.1850 & 0.4832 & 0.3110  \\
                                     & \textbf{GAR}         & 0.1331                        & 0.1840                             & 0.2126                        & \underline {0.2195}                       & 0.2380                        & 0.2033                        & { 0.1384} & { 0.1027}      & { 0.2285} & { 0.1328}      & { 0.2196} & { 0.1326} & 0.2594                        & 0.2042                             & \underline {0.3548}                  & \underline {0.2499}                       & 0.4062                        & 0.2514                        \\
                                     & \textbf{GoRec}       & 0.0807                        & 0.1207                             & \underline {0.2162}                  & 0.2166                             & 0.1171                        & 0.0994                        & 0.0808                        & 0.0598                             & 0.2398                        & 0.1345                             & 0.1114                        & 0.0612                        & 0.1434                        & 0.1169                             & 0.2372                        & 0.1705                             & 0.2343                        & 0.1438                        \\
                                     & \textbf{CLCRec}      & 0.1294                        & 0.1894                             & 0.1553                        & 0.1515                             & 0.2805                        & 0.2465                        & \underline {0.1714}                  & \underline {0.1278}                       & \underline {0.2790}                  & \underline {0.1610}                       & 0.2611                        & 0.1591                        & \underline {0.2697}                        & \underline {0.2208}                             & 0.3429                        & 0.2451                             & 0.4832                        & 0.3110                        \\
                                     & \textbf{ALDI}        & \underline {0.1451}                  & \underline {0.2065}                       & 0.2142                        & 0.2137                             & \underline{0.2805}                       & \underline {0.2465}                  & 0.1607                        & 0.1217                             & 0.2570                        & 0.1472                             & 0.2611                        & 0.1591                        & 0.2622                        & 0.2199                             & 0.2914                        & 0.1986                             & \underline {0.4832}                        & \underline {0.3110 }                      \\
                                     & \textbf{PAD-CLRec} &  0.1359 & 0.1966 & 0.2046 & 0.2002 & 0.2749 & 0.2417& 0.1799 & 0.1342 & 0.2912 & 0.1711 & \underline{0.2643} & \underline{0.1597} & 0.2533 & 0.1987 & 0.3134 & 0.1876 & 0.4620 & 0.2971 \\
                                     \cline{2-20}
\rowcolor{cyan!15}                   &  \textbf{DiffCold (Ours)}    & \textbf{0.1596}                        & \textbf{0.2174}                             & \textbf{0.2341}                      & \textbf{0.2262}                             & \textbf{0.2987}                        & \textbf{0.2661}                       & \textbf{0.2027}                        & \textbf{0.1517}                             & \textbf{0.3322}                        & \textbf{0.1926}                             & \textbf{0.2737}                        & \textbf{0.1671}                        & \textbf{0.2847}                        & \textbf{0.2357}                             & \textbf{0.3837}                        & \textbf{0.2704}                             & \textbf{0.4968}                        & \textbf{0.3256}                       \\
\rowcolor{cyan!15}\multirow{-11}{*}{\textbf{LightGCN}} &  \textbf{Improv.} & 9.99\%                        & 5.28\%                              & 8.28\%                         & 3.05\%                              & 6.49\%                         & 7.95\%                         & 18.26\%                        & 18.70\%                             & 19.07\%                        & 19.63\%                             & 3.56\%                         & 4.63\%                         & 5.56\%                         & 6.75\%                              & 8.15\%                         & 8.20\%                              & 2.81\%                         & 4.69\%                         \\ \hline \hline
\rowcolor{yellow!10}  & \textbf{Backbone}    & 0.1230                        & 0.1776                             & 0.0000                        & 0.0000                             & 0.2721                        & 0.2382                        & 0.1345                        & 0.1160                             & 0.0000                        & 0.0000                             & 0.2922                        & 0.1852                        & 0.2051                        & 0.1841                             & 0.0000                        & 0.0000                             & 0.4879                        & 0.3179  \\ \cline{2-20}
                                     & \textbf{DropoutNet}  & 0.0886                        & 0.1377                             & 0.1459                        & 0.1443                             & 0.1882                        & 0.1754                        & 0.0534                        & 0.0413                             & 0.1817                        & 0.1002                             & 0.1126                        & 0.0639                        & 0.1259                        & 0.1072                             & 0.1525                        & 0.1142                             & 0.2584                        & 0.1677                        \\
                                     & \textbf{Heater}      & 0.1230                        & 0.1763                             & 0.1586                        & 0.1545                             & 0.2713                        & 0.2358                        & 0.1220                        & 0.1042                             & 0.1631                        & 0.1046                             & 0.2647                        & 0.1656                        & 0.2149                        & 0.1880                             & 0.1781                        & 0.1437                             & 0.4812                        & 0.3124                        \\
                                     & \textbf{MetaEmb}  & 0.1263  &  0.1784 & 0.1677 & 0.1631 & 0.2721 & 0.2382 & 0.1399 & 0.1108 & 0.1873 & 0.1147 & 0.2922 & 0.1852 & 0.2354 & 0.1911 & 0.3036 & 0.2176 & 0.4879 & 0.3179 \\
                                     & \textbf{GAR}         & 0.1298                        & 0.1805                             & 0.1862                        & 0.1826                             & 0.2542                        & 0.2237                        & 0.1580                        & 0.1188                             & 0.2084                        & 0.1232                             & 0.2910                        & 0.1846                        & \underline {0.2598}                  & 0.2072                             & \underline {0.3499}                  & \underline {0.2529}                       & 0.4115                        & 0.2588                        \\
                                     & \textbf{GoRec}       & 0.0778                        & 0.1235                             & 0.1591                        & 0.1703                             & 0.1444                        & 0.1357                        & 0.0903                        & 0.0649                             & 0.2436                        & 0.1390                             & 0.1229                        & 0.0676                        & 0.1772                        & 0.1584                             & 0.2141                        & 0.1688                             & 0.2465                        & 0.1872                        \\
                                     & \textbf{CLCRec}      & \underline {0.1504}                  & \underline {0.2119}                       & \underline {0.2029}                  & \underline {0.2003}                       & 0.2721                        & 0.2382                        & 0.1532                        & 0.1203                             & \underline {0.2948}                  & \underline {0.1751}                       & 0.2922                        & 0.1852                        & \multicolumn{1}{l}{0.2384}    & \multicolumn{1}{l}{0.2005}         & \multicolumn{1}{l}{0.3142}    & \multicolumn{1}{l}{0.2306}         & \multicolumn{1}{l}{0.4879}    & \multicolumn{1}{l}{0.3179}    \\
                                     & \textbf{ALDI}        & 0.1313                        & 0.1847                             & 0.1413                        & 0.1425                             & \underline {0.2721 }                      & \underline {0.2382}                        &  0.1709                  & 0.1302                     & 0.2555                        & 0.1465                             & 0.2922                        & 0.1852                  & 0.2596                        & \underline {0.2198}                       & 0.2798                        & 0.1961                             & \underline {0.4879}                        & \underline {0.3179}
                                     \\
                  & \textbf{PAD-CLRec} & 0.1427 & 0.1977 & 0.1898 & 0.1841 & 0.2654 & 0.2319 & \underline {0.1817}  & \underline {0.1443} & 0.2871 & 0.1642 & \underline{0.2971} &  \underline {0.1879} & 0.2543 & 0.2004 & 0.3173 & 0.2344 & 0.4679 & 0.2903\\   \cline{2-20}
\rowcolor{cyan!15} &   \textbf{DiffCold (Ours)}    & \textbf{0.1620}                        & \textbf{0.2232}                             & \textbf{0.2329}                        & \textbf{0.2257}                             & \textbf{0.2955}                        & \textbf{0.2605}                        & \textbf{0.2246}                        & \textbf{0.1692}                             & \textbf{0.3439}                        & \textbf{0.2020}                             & \textbf{0.3096}                        & \textbf{0.1949}                        & \textbf{0.2888}                        & \textbf{0.2359 }                            & \textbf{0.3916}                        & \textbf{0.2668}                             & \textbf{0.4956}                        & \textbf{0.3245}                        \\
\rowcolor{cyan!15}\multirow{-11}{*}{\textbf{SimGCL}}   &  \textbf{Improv. (\%)} & 7.71\%                         & 5.33\%                              & 14.79\%                        & 12.68\%                             & 4.60\%                         & 4.08\%                         & 23.61\%                        & 17.26\%                             & 16.66\%                        & 15.36\%                             & 4.20\%                         & 3.72\%                         & 11.16\%                        & 7.32\%                              & 11.92\%                        & 1.54\%                              & 1.58\%                         & 2.08\%                         \\ \hline

\end{tabular}
}\label{Table:Overall}
\end{table}

We conduct comprehensive experiments to evaluate the performance of DiffCold and answer the following questions:
\begin{itemize}
    \item \textbf{RQ1}: How does DiffCold perform as compared to the various baselines in cold-start item recommendation?
    \item \textbf{RQ2}: How can DiffCold solve the starting-point problem?
    \item \textbf{RQ3}: Does DiffCold address the distribution-consistency problem by aligning the item embeddings of warm and cold items?
    \item \textbf{RQ4}: What is the effect of different components and loss terms within DiffCold?
    \item \textbf{RQ5}: How do the hyperparameters affect the performance of DiffCold?
    \item \textbf{RQ6}: How does the efficiency of DiffCold compare with other methods?

\end{itemize}

\subsection{Experiment Setups}
\textbf{Datasets and Evaluation Metrics}
We conducted experiments on three open-source datasets: Movielens, Citeulike, and Xing. Dataset statistics are shown in Table \ref{Table:Statistics of datasets}. Item content features are encoded with different dimensions following previous works~\cite{Heater,ALDI}. For each dataset, 20\% of items are treated as cold-start, with their interactions split equally into validation and test sets (1:1). The remaining 80\% of items’ interactions are divided into training, validation, and test sets in an 8:1:1 ratio. Recommendation performance is evaluated using Recall@K and NDCG@K, with K=20.

\noindent \textbf{Backbones and Baselines}
We use three backbone models (\ie, MF~\cite{rendle2012bpr}, LightGCN~\cite{he2020lightgcn} and SimGCL~\cite{SimGCL}) to serve as the initialization for warm items. We compare DiffCold with the following methods: (1) Dropout-based methods: DropoutNet~\cite{DropoutNet}, Heater~\cite{Heater}; (2) Generative methods: MetaEmb~\cite{MetaEmb}, GAR~\cite{GAR} and GoRec~\cite{GoRec}; (3) Alignment-based methods: CLCRec~\cite{CLCRec}, ALDI~\cite{ALDI}, PAD-CLRec~\cite{PAD-CLRec}.

\noindent \textbf{Implementation Details}
All models are implemented using the comprehensive open-source framework ColdRec\footnote{https://github.com/YuanchenBei/ColdRec}. Each model undergoes a corresponding hyperparameter search described in original papers to ensure fair comparison.
We employ $64$-dimensional embeddings and optimize the parameters with the Adam optimizer ($lr = 0.001$), a batch size of $2048$, and an $\ell_{2}$ regularization weight of $0.0001$.
The Retrieval-Enhanced Aggregator is configured with $k = 10$, and the number of diffusion steps is set to $T = 20$.
The influence of the hyperparameter in Eq.~\eqref{Eq:overall} is examined in Section~\ref{sec:hyperparameter}.






\subsection{Overall Performance Comparison (RQ1)}
We compare DiffCold with other baselines based on different backbones. The results are reported in Table~\ref{Table:Overall}, from which we can observe: \textbf{(i)} Both dropout-based methods
   and generative methods exhibit the seesaw dilemma, whereby optimizing the performance for cold items leads to a significant decline in the performance for warm items. Although alignment-based methods do not update the embeddings of warm items, their overall and cold recommendation performance remains suboptimal. \textbf{(ii)} DiffCold consistently outperforms all baselines in terms of overall, cold, and warm recommendation across all datasets and backbones, effectively addressing the seesaw problem. This advantage is attributed to two key modules in DiffCold: the Retrieval-enhanced Aggregator, which resolves the starting-point problem, and the Simulation-based Representation Alignment, which further ensures distributional consistency.

\subsection{Starting-Point Problem (RQ2)}

\begin{figure}
    \centering
\includegraphics[width=1\linewidth]{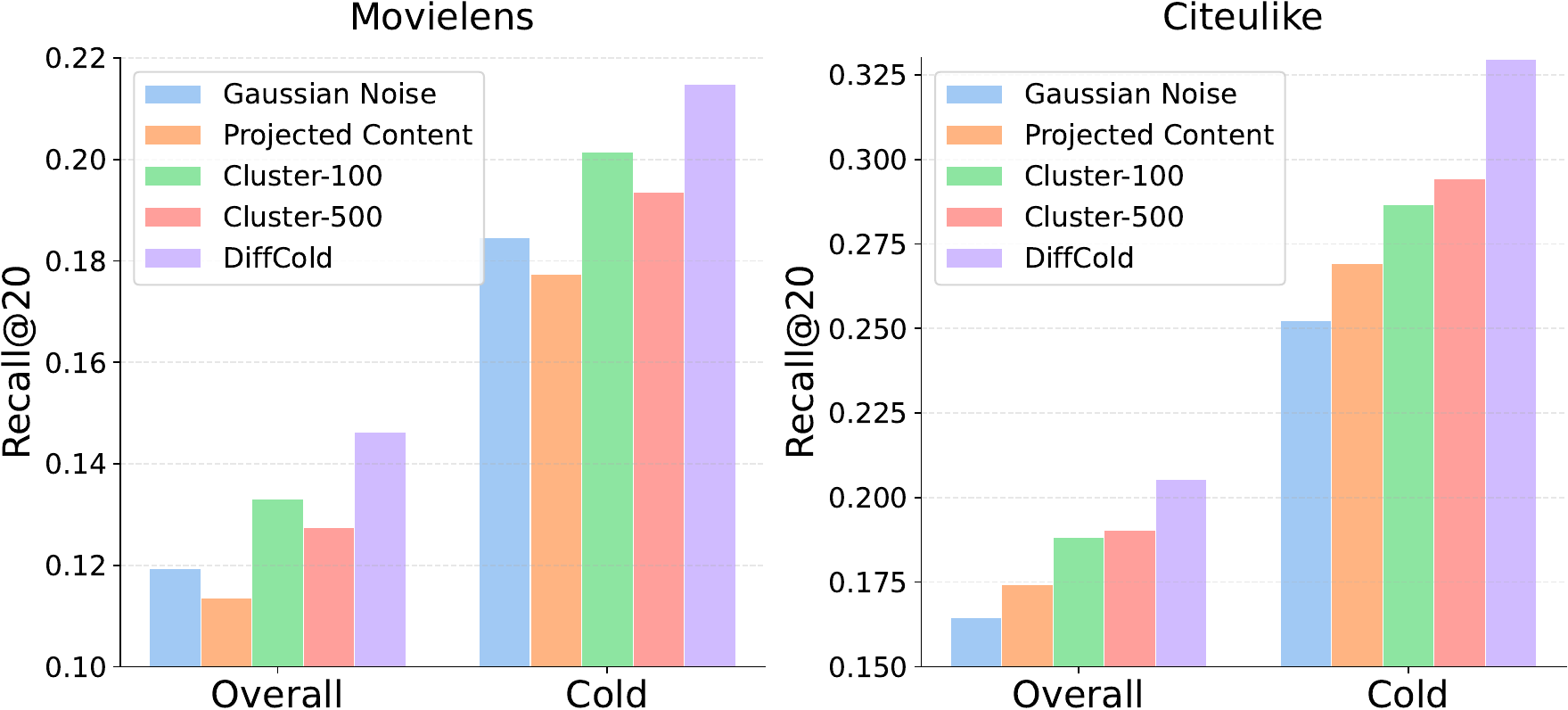}
    \caption{The comparison between different starting representation methods based on MF backbone.}
    \label{fig:RQ2}
\end{figure}
To further verify the impact of different starting representations, we replace the starting point with the following variants:
(1) \textbf{Gaussian Noise};
(2) \textbf{Projected Content Features}: Content feature is projected through a linear layer for dimensionality alignment;
(3) \textbf{Cluster-$\textbf{100}$}: We apply KNN (100 centroids) to content features, using each cluster's average representation.
(4) \textbf{Cluster-$\textbf{500}$}: KNN with 500 centroids. The result in Figure \ref{fig:RQ2} shows that relying exclusively on Gaussian noise severely impairs the personalization information of cold items. Although clustering methods partly enhance the recommendation performance, they are highly dependent on hyperparameters and the dataset's distribution. The Retrieval-enhanced Aggregator achieves the optimal performance.

\begin{table}[ht]
\centering
\caption{The quantitative analysis of the distribution of warm and cold items. `Wass. Dist.' refers to `Wasserstein Distance'. `MMD' refers to `Maximum Mean Discrepancy'.}
\begin{tabular}{c|cccc|cccc}
\hline
\multirow{2}{*}{\textbf{Methods}}
& \multicolumn{4}{c|}{\cellcolor{gray!20}\textbf{Movielens}}
& \multicolumn{4}{c}{\cellcolor{gray!20}\textbf{Citeulike}} \\ \cline{2-9}
 & \cellcolor{gray!20}\begin{tabular}[c]{@{}c@{}} Intra-\\ Cold($\downarrow$)\end{tabular}
 & \cellcolor{gray!20}\begin{tabular}[c]{@{}c@{}}Intra-\\ Warm ($\downarrow$)\end{tabular}
 & \cellcolor{gray!20}\begin{tabular}[c]{@{}c@{}}Wass\\ Dist ($\downarrow$)\end{tabular}
 & \cellcolor{gray!20}MMD ($\downarrow$)
 & \cellcolor{gray!20}\begin{tabular}[c]{@{}c@{}}Intra-\\ Cold ($\downarrow$)\end{tabular}
 & \cellcolor{gray!20}\begin{tabular}[c]{@{}c@{}}Intra-\\ Warm ($\downarrow$)\end{tabular}
 & \cellcolor{gray!20}\begin{tabular}[c]{@{}c@{}}Wass\\ Dist ($\downarrow$)\end{tabular}
 & \cellcolor{gray!20}MMD ($\downarrow$) \\ \hline
\textbf{Heater}   & .9740 & .0404 & 6.435 & .6354 & .7021 & .0185 & 2.352 & .5300 \\
\textbf{GAR}      & .4712 & .0605 & \textbf{2.655} & .0983 & .1800 & .0013 & 1.854 & .0941 \\
\textbf{ALDI}     & .6512 & .0123 & 11.24 & .0945 & .0818 & .0029 & 3.614 & \textbf{.0244} \\
\rowcolor{cyan!10} \textbf{DiffCold} & \textbf{.0437} & \textbf{.0057} & 4.167 & \textbf{.0549} & \textbf{.0112} & \textbf{.0010} & \textbf{1.815} & .0373 \\ \hline
\end{tabular}%
\label{Table:quantitative analysis}
\end{table}

\subsection{Distribution-Consistency Problem (RQ3)}
\begin{figure}[t]
    \centering
\includegraphics[width=0.8\linewidth]{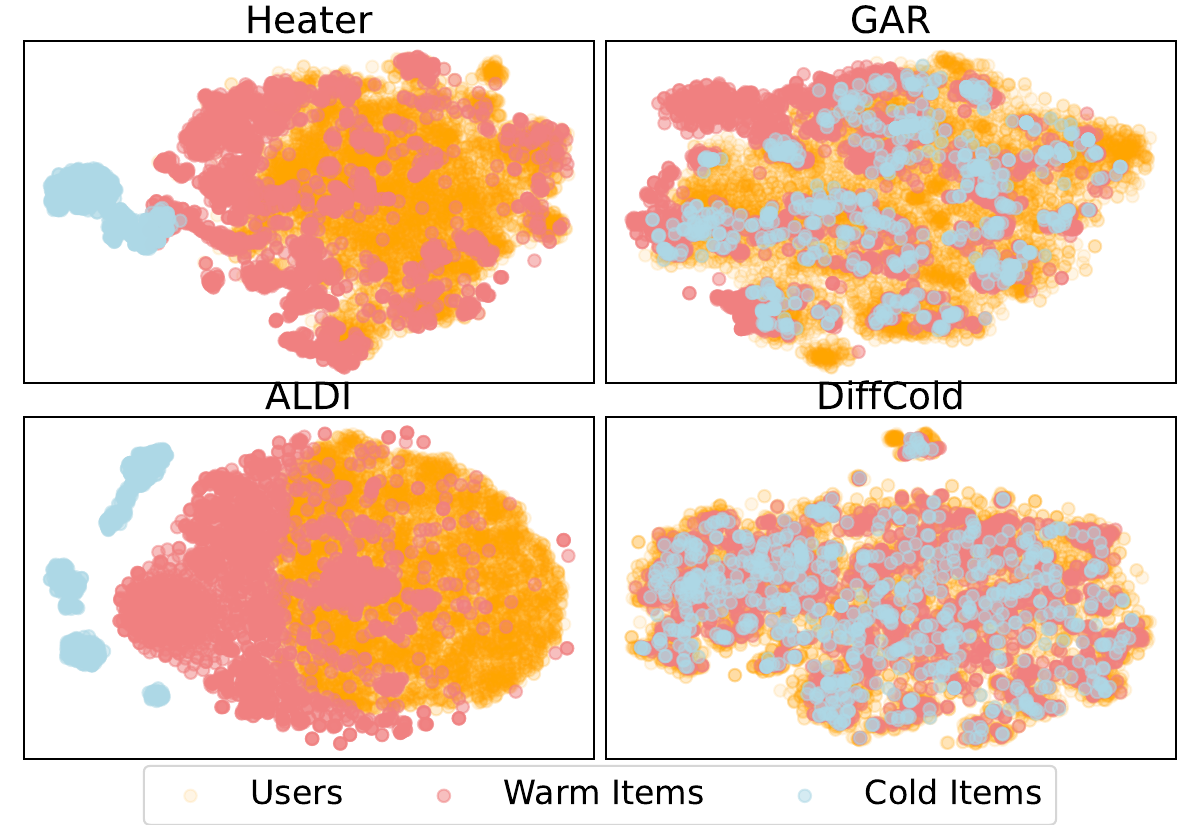}
    \caption{The distribution of learned ID representations of users, warm and cold items on Movielens.}
    \label{fig:vis_distribution}
\end{figure}

We evaluate the distribution divergence between warm and cold items using the following metrics: (1) \textbf{Intra-Cold/Intra-Warm Similarity} computes the average cosine similarity within cold and warm representations respectively; (2) \textbf{Wasserstein Distance}~\cite{WassersteinDis} computes the minimum cost required to transform the distribution of cold items into warm items; (3) \textbf{Maximum Mean Discrepancy}~\cite{MMD} measures the distribution divergence based on kernel methods.
Results in Table \ref{Table:quantitative analysis} show DiffCold achieves the lowest Intra-Cold/Intra-Warm values, indicating stronger item discrimination. The smallest difference between Intra-Cold/Intra-Warm implies the highest consistency in representation distributions. DiffCold also attains optimal/near-optimal performance in Wasserstein/MMD measures. Figure \ref{fig:vis_distribution} further shows the ID distributions. Heater and ALDI demonstrate a clear separation between cold and warm items, with user representations clustered near the warm items. GAR partially reduces inconsistencies but leaves small clusters under-optimized. DiffCold demonstrates superior distribution alignment between cold/warm items, as well as better user-item representation integration.

\subsection{Ablation Study (RQ4)}\label{sec:ablation}

\begin{table}[ht]
\centering
\caption{The ablation study (Recall@20) about DiffCold.}
\label{Table:AblationStudy}
\begin{tabular}{c|c|ccc|ccc}
\hline
\multirow[c]{2}{*}{\textbf{Backbone}} &
\multirow[c]{2}{*}{\textbf{Variants}} &
\multicolumn{3}{c|}{\textbf{Movielens}} &
\multicolumn{3}{c}{\textbf{Citeulike}} \\
\cline{3-8}
&  & \textbf{Overall} & \textbf{Cold} & \textbf{Warm} & \textbf{Overall} & \textbf{Cold} & \textbf{Warm} \\
\hline

\multirow{6}{*}{\textbf{MF}}
& W/O Aggregator    & 0.1194 & 0.1846 & 0.2583 & 0.1645 & 0.2524 & 0.2783 \\
& W/O Sim-Align     & 0.1404 & 0.1953 & 0.2589 & 0.1933 & 0.2921 & 0.2739 \\
& W/O BPR-Cold      & 0.1353 & 0.2104 & 0.2457 & 0.1028 & 0.2278 & 0.2695 \\
& W/O BPR-Warm      & 0.1289 & 0.2042 & 0.2222 & 0.1361 & 0.2891 & 0.2283 \\
& W/O Warm-Aug      & 0.1365 & 0.2124 & 0.2396 & 0.1786 & 0.3024 & 0.2497 \\
\rowcolor{cyan!15}& \textbf{DiffCold} & \textbf{0.1463} & \textbf{0.2148} & \textbf{0.2593} & \textbf{0.2054} & \textbf{0.3295} & \textbf{0.2873} \\
\hline

\multirow{6}{*}{\textbf{LightGCN}}
& W/O Aggregator    & 0.1369 & 0.2065 & 0.2836 & 0.1585 & 0.2582 & 0.2734 \\
& W/O Sim-Align     & 0.1450 & 0.2040 & 0.2771 & 0.1537 & 0.2511 & 0.2709 \\
& W/O BPR-Cold      & 0.1505 & 0.2136 & 0.2844 & 0.1495 & 0.2244 & \textbf{0.2819} \\
& W/O BPR-Warm      & 0.1491 & 0.2267 & 0.2734 & 0.1489 & 0.2531 & 0.2639 \\
& W/O Warm-Aug      & 0.1525 & 0.2251 & 0.2768 & 0.1882 & 0.3172 & 0.2531 \\
\rowcolor{cyan!15}&  \textbf{DiffCold} & \textbf{0.1596} & \textbf{0.2341} & \textbf{0.2987} & \textbf{0.2027} & \textbf{0.3322} & 0.2737 \\
\hline
\end{tabular}%
\end{table}

We analyze how each of the proposed components affects the performance of DiffCold:
(1) $\textbf{W/O Aggregator}$: The Aggregator is replaced with Gaussian noise;
(2) $\textbf{W/O Sim-Align}$: $ L_{\text{sim-align}}$ is removed ; (3) \textbf{W/O BPR-Cold}: $L_{\text{BPR-Cold}} $ is removed;
(4) \textbf{W/O BPR-Warm}: $L_{\text{bpr-warm}} $ is removed;
(5) {\textbf{W/O Warm-Aug}}: $L_{\text{warm-aug}}$ is removed.
With the result illustrated in Table \ref{Table:AblationStudy}, all components contribute to the overall performance of DiffCold. Specifically, `W/O Aggregator', `W/O Sim-Align', and `W/O BPR-Cold' result in a more significant decrease in the cold item recommendation, whereas `W/O BPR-Warm' and `W/O Warm-Aug' tend to primarily affect warm items.


\subsection{Hyperparamter Study  (RQ5)}\label{sec:hyperparameter}
\begin{figure}[h]
    \centering
    \includegraphics[width=1\linewidth]{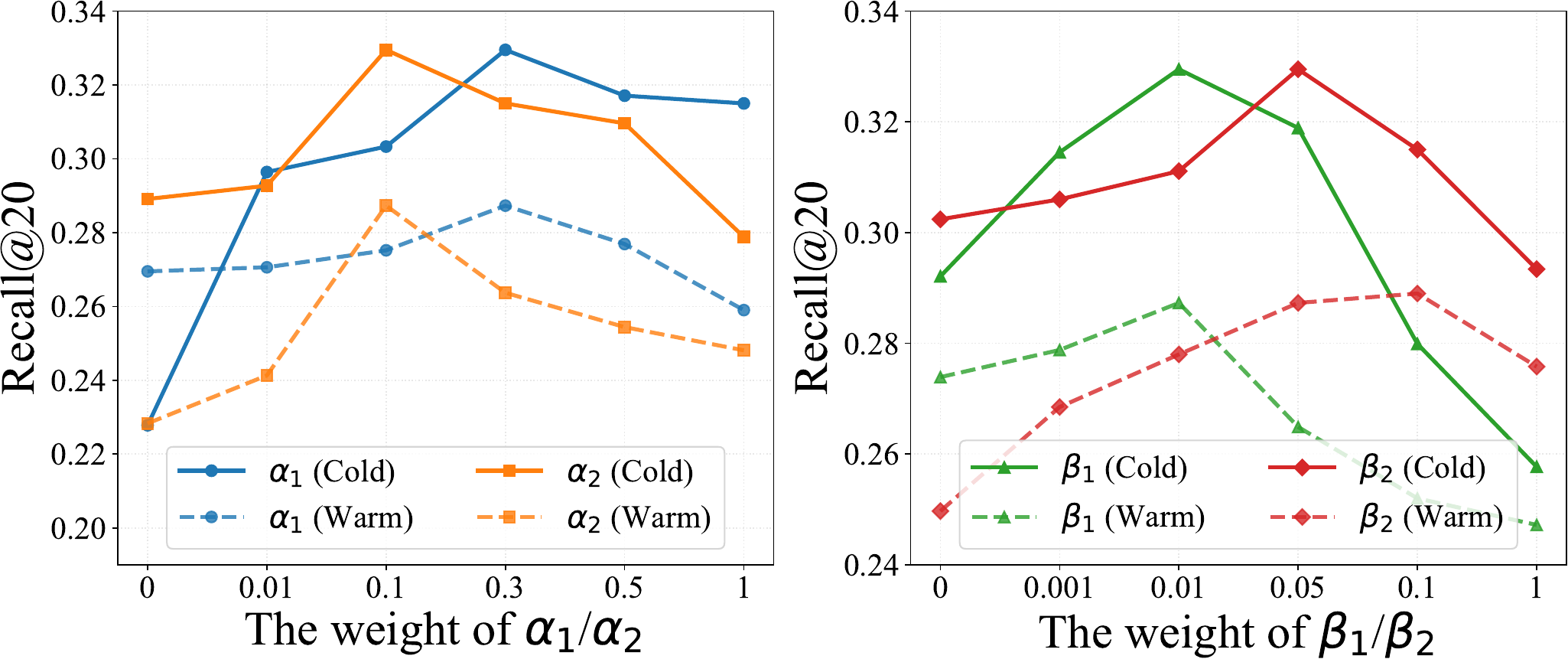}
    \caption{The hyperparamter result of $\alpha_1$, $\alpha_2$ and $\beta_1$, $\beta_2$ on Citeulike based on MF.}
    \label{fig:hyper-parameter}
\end{figure}

We first investigate four key hyperparameters in DiffCold: $\alpha_1$, $\alpha_2$, $\beta_1$, and $\beta_2$ as defined in Equation \eqref{Eq:overall}. Figure \ref{fig:hyper-parameter} illustrates how these hyperparameters influence the recommendation performance for both cold and warm items. In general, when these parameters vary within reasonable ranges, the performance exhibits only modest fluctuations, demonstrating the model's robustness.

We further investigate the impact of diffusion steps and the scheduler on the DiffCold's recommendation performance. The experimental results in Table \ref{tab:appendix_diffusion_step} indicate that, within the generation pipeline of DiffCold, longer diffusion steps significantly increase both the training and inference time of the model. We adopt a linear scheduler and 20 inference steps for optimal performance.

\begin{table}[ht]
\centering
\caption{The experimental results for DiffCold based on MF in Movielens under different combinations of diffusion steps and schedulers.}
\begin{tabular}{c|cccc}
\hline
\rowcolor{gray!20}
\textbf{Recall@20}       & \textbf{Overall} & \textbf{Cold}   & \textbf{Warm}   & \textbf{Time(s)/epoch} \\ \hline
\textbf{Linear, 5-step}    & 0.1429 & 0.2071 & 0.2494 & 10.99 \\
\textbf{Linear, 10-step}   & 0.1431 & 0.2099 & 0.2529 & 12.34 \\
\rowcolor{cyan!15}
\textbf{Linear, 20-step} & \textbf{0.1463}  & \textbf{0.2148} & \textbf{0.2593} & \textbf{14.04}      \\
\textbf{Linear, 100-step}  & 0.1443 & 0.2081 & 0.2586 & 20.39 \\
\textbf{Cosine, 20-step}   & 0.1143 & 0.1850 & 0.2225 & 15.71 \\
\textbf{Binomial, 20-step} & 0.1211 & 0.1972 & 0.2244 & 16.04 \\ \hline
\end{tabular}%
\label{tab:appendix_diffusion_step}
\end{table}


\subsection{The Efficiency Analysis (RQ6)}
\begin{table}[ht]
\centering
\caption{The efficiency analysis of different models. The experiments are conducted on a Tesla V100 GPU card with 32 GB memory.}
\begin{tabular}{c|cccc}
\hline
\rowcolor{gray!20}
\textbf{Dataset} & \textbf{Models} & \textbf{\begin{tabular}[c]{@{}c@{}}run memory \\ (MB)\end{tabular}} & \textbf{\# epochs} & \textbf{time/epoch(s)} \\ \hline
\multirow{5}{*}{\textbf{Movielens}} & Heater   & 977  & 36  & 14.55 \\
                                    & GAR      & 953  & 113 & 15.76 \\
                                    & CLCRec   & 1741 & 16  & 210   \\
                                    & ALDI     & 985  & 28  & 48.40 \\
\rowcolor{cyan!15}                  & DiffCold & 1199 & 21  & 14.04 \\ \hline
\multirow{5}{*}{\textbf{Citeulike}} & Heater   & 1067 & 140 & 5.64  \\
                                    & GAR      & 1009 & 92  & 5.99  \\
                                    & CLCRec   & 1791 & 28  & 50.64 \\
                                    & ALDI     & 1019 & 51  & 6.44  \\
\rowcolor{cyan!15}                  & DiffCold & 1278 & 36  & 5.88  \\ \hline
\end{tabular}%
\label{Table:Efficiency}
\end{table}

In this section, we focus on analyzing the differences in efficiency across various models, primarily examining metrics such as memory usage during runtime, computational duration, and convergence speed. The experiment results are presented in Table \ref{Table:Efficiency}. CLCRec~\cite{CLCRec} significantly occupies larger memory consumption and computational resources.
In contrast, DiffCold achieves a better balance compared to other models, attaining consistent improvements in warm, cold, and overall recommendations while occupying nearly the same amount of memory and computational resources as other models.

\section{Conclusion}
This study identifies the seesaw dilemma in cold-start recommendation, where distinct representation distributions hinder simultaneous performance improvement for cold and warm items. To address this, we propose DiffCold, a generative model with two key components: Retrieval-enhanced Aggregator for inference initialization and Simulation-based Representation Alignment for generating synthetic cold representations to ensure distribution consistency. Experiments and ablation studies validate DiffCold's effectiveness and the robustness of generated representations.

\begin{credits}
\subsubsection{\ackname}
This paper is supported by National Natural Science Foundation of China (624B2096, 72595872, 72542012, 62322603).

\subsubsection{\discintname}
The authors have no competing interests to declare that are relevant to the content of this article.
\end{credits}

\bibliographystyle{splncs04}
\bibliography{main}

\begin{thebibliography}{10}
\providecommand{\url}[1]{\texttt{#1}}
\providecommand{\urlprefix}{URL }
\providecommand{\doi}[1]{https://doi.org/#1}

\bibitem{GoRec}
Bai, H., Hou, M., Wu, L., Yang, Y., Zhang, K., Hong, R., Wang, M.: Gorec: A
  generative cold-start recommendation framework. In: Proceedings of the 31st
  ACM International Conference on Multimedia. p. 1004–1012. MM '23 (2023)

\bibitem{GAR}
Chen, H., Wang, Z., Huang, F., Huang, X., Xu, Y., Lin, Y., He, P., Li, Z.:
  Generative adversarial framework for cold-start item recommendation. In:
  Proceedings of the 45th International ACM SIGIR Conference on Research and
  Development in Information Retrieval. p. 2565–2571. SIGIR '22 (2022)

\bibitem{CaDiRec}
Cui, Z., Wu, H., He, B., Cheng, J., Ma, C.: Context matters: Enhancing
  sequential recommendation with context-aware diffusion-based contrastive
  learning. In: Proceedings of the 33rd ACM International Conference on
  Information and Knowledge Management. pp. 404--414. CIKM '24 (2024)

\bibitem{KZScholarFints}
Du, K., Liu, J., Zhang, K., Jiao, W., Lu, Y., Jin, J., Liu, W., Yu, Y., Zhang,
  W.: {Fints}: Efficient inference-time personalization for {LLMs} with
  fine-grained instance-tailored steering (2025)

\bibitem{MTPR}
Du, X., Wang, X., He, X., Li, Z., Tang, J., Chua, T.S.: How to learn item
  representation for cold-start multimedia recommendation? In: Proceedings of
  the 28th ACM International Conference on Multimedia. p. 3469–3477. MM '20
  (2020)

\bibitem{KZScholarCodeApex}
Fu, L., Chai, H., Luo, S., Du, K., Zhang, W., Fan, L., Lei, J., Rui, R., Lin,
  J., Fang, Y., Liu, Y., Wang, J., Qi, S., Zhang, K., Zhang, W., Yu, Y.:
  {CodeApex}: A bilingual programming evaluation benchmark for large language
  models (2024)

\bibitem{DiffuSeq}
Gong, S., Li, M., Feng, J., Wu, Z., Kong, L.: Diffuseq: Sequence to sequence
  text generation with diffusion models (2023)

\bibitem{KZScholarSWECycle}
Guan, H., Fu, L., Zhang, S., Zhu, Y., Zhang, K., Qiu, L., Cai, X., Cao, X.,
  Liu, W., Zhang, W., Yu, Y.: {SWE-Cycle}: Benchmarking code agents across the
  complete issue resolution cycle (2026)

\bibitem{he2020lightgcn}
He, X., Deng, K., Wang, X., Li, Y., Zhang, Y., Wang, M.: Lightgcn: Simplifying
  and powering graph convolution network for recommendation. In: Proceedings of
  the 43rd International ACM SIGIR Conference on Research and Development in
  Information Retrieval. pp. 639--648. SIGIR'20 (2020)

\bibitem{DDPM}
Ho, J., Jain, A., Abbeel, P.: Denoising diffusion probabilistic models.
  Advances in neural information processing systems  \textbf{33},  6840--6851
  (2020)

\bibitem{ALDI}
Huang, F., Wang, Z., Huang, X., Qian, Y., Li, Z., Chen, H.: Aligning
  distillation for cold-start item recommendation. In: Proceedings of the 46th
  International ACM SIGIR Conference on Research and Development in Information
  Retrieval. p. 1147–1157. SIGIR '23 (2023)

\bibitem{KZScholarClickPrompt}
Lin, J., Chen, B., Wang, H., Xi, Y., Qu, Y., Dai, X., Zhang, K., Tang, R., Yu,
  Y., Zhang, W.: {ClickPrompt}: {CTR} models are strong prompt generators for
  adapting language models to {CTR} prediction. In: Proceedings of the ACM Web
  Conference 2024. pp. 3319--3330 (2024)

\bibitem{DiffRec_survey}
Lin, J., Liu, J., Zhu, J., Xi, Y., Liu, C., Zhang, Y., Yu, Y., Zhang, W.: A
  survey on diffusion models for recommender systems. arXiv preprint
  arXiv:2409.05033  (2024)

\bibitem{DiffuSAR}
Liu, Q., Yan, F., Zhao, X., Du, Z., Guo, H., Tang, R., Tian, F.: Diffusion
  augmentation for sequential recommendation. In: Proceedings of the 32nd ACM
  International Conference on Information and Knowledge Management. p.
  1576–1586. CIKM '23 (2023)

\bibitem{KZScholarAlignRec}
Liu, Y., Zhang, K., Ren, X., Huang, Y., Jin, J., Qin, Y., Su, R., Xu, R., Yu,
  Y., Zhang, W.: {AlignRec}: Aligning and training in multimodal
  recommendations. In: Proceedings of the 33rd ACM International Conference on
  Information and Knowledge Management. pp. 1503--1512 (2024)

\bibitem{SeeDRec}
Ma, H., Xie, R., Meng, L., Yang, Y., Sun, X., Kang, Z.: Seedrec: sememe-based
  diffusion for sequential recommendation. In: Proceedings of the Thirty-Third
  International Joint Conference on Artificial Intelligence. pp. 2270--2278.
  IJCAI '24 (2024)

\bibitem{DeepMusic}
Oord, A.v.d., Dieleman, S., Schrauwen, B.: Deep content-based music
  recommendation. In: Advances in Neural Information Processing Systems 26. p.
  2643–2651. NIPS'13 (2013)

\bibitem{MetaEmb}
Pan, F., Li, S., Ao, X., Tang, P., He, Q.: Warm up cold-start advertisements:
  Improving ctr predictions via learning to learn id embeddings. In:
  Proceedings of the 42nd International ACM SIGIR Conference on Research and
  Development in Information Retrieval. p. 695–704. SIGIR'19 (2019)

\bibitem{rendle2012bpr}
Rendle, S., Freudenthaler, C., Gantner, Z., Schmidt-Thieme, L.: Bpr: Bayesian
  personalized ranking from implicit feedback. arXiv preprint arXiv:1205.2618
  (2012)

\bibitem{KZScholarAutoGraphRec}
Shan, R., Lin, J., Zhu, C., Chen, B., Zhu, M., Zhang, K., Zhu, J., Tang, R.,
  Yu, Y., Zhang, W.: An automatic graph construction framework based on large
  language models for recommendation. In: Proceedings of the 31st ACM SIGKDD
  Conference on Knowledge Discovery and Data Mining. pp. 4806--4817 (2025)

\bibitem{MMD}
Smola, A.J., Gretton, A., Borgwardt, K.M.: Maximum mean discrepancy. In:
  Proceedings of the 13th International Conference on Neural Information
  Processing. vol.~6 (2006)

\bibitem{DDIM}
Song, J., Meng, C., Ermon, S.: Denoising diffusion implicit models. arXiv
  preprint arXiv:2010.02502  (2020)

\bibitem{WassersteinDis}
VM, P.: Statistical aspects of wasserstein distances. Annual review of
  statistics and its application  \textbf{6}(1),  405--431 (2019)

\bibitem{DropoutNet}
Volkovs, M., Yu, G., Poutanen, T.: Dropoutnet: addressing cold start in
  recommender systems. In: Advances in Neural Information Processing Systems
  30. p. 4964–4973. NIPS'17 (2017)

\bibitem{PAD-CLRec}
Wang, W., Liu, B., Shan, L., Sun, C., Chen, B., Guan, J.: Preference aware dual
  contrastive learning for item cold-start recommendation. In: Proceedings of
  the AAAI Conference on Artificial Intelligence. AAAI'24/IAAI'24/EAAI'24,
  vol.~38, pp. 9125--9132 (2024)

\bibitem{DiffRec}
Wang, W., Xu, Y., Feng, F., Lin, X., He, X., Chua, T.S.: Diffusion recommender
  model. In: Proceedings of the 46th International ACM SIGIR Conference on
  Research and Development in Information Retrieval. p. 832–841. SIGIR '23
  (2023)

\bibitem{CLCRec}
Wei, Y., Wang, X., Li, Q., Nie, L., Li, Y., Li, X., Chua, T.S.: Contrastive
  learning for cold-start recommendation. In: Proceedings of the 29th ACM
  International Conference on Multimedia. p. 5382–5390. MM '21 (2021)

\bibitem{DreamRec}
Yang, Z., Wu, J., Wang, Z., Yuan, Y., Wang, X., He, X.: Generate what you
  prefer: reshaping sequential recommendation via guided diffusion. In:
  Advances in Neural Information Processing Systems 36. NIPS '23 (2023)

\bibitem{SimGCL}
Yu, J., Yin, H., Xia, X., Chen, T., Cui, L., Nguyen, Q.V.H.: Are graph
  augmentations necessary? simple graph contrastive learning for
  recommendation. In: Proceedings of the 45th International ACM SIGIR
  Conference on Research and Development in Information Retrieval. pp.
  1294--1303 (2022)

\bibitem{KZScholarDP3}
Ze, Y., Zhang, G., Zhang, K., Hu, C., Wang, M., Xu, H.: {3D} diffusion policy:
  Generalizable visuomotor policy learning via simple {3D} representations
  (2024)

\bibitem{KZScholarLoopTool}
Zhang, K., Jiao, W., Du, K., Lu, Y., Liu, W., Zhang, W., Yu, Y.: {LoopTool}:
  Closing the data-training loop for robust {LLM} tool calls. In: Proceedings
  of the 64th Annual Meeting of the Association for Computational Linguistics
  (2025)

\bibitem{KZScholarTokenCrossing}
Zhang, K., Jin, J., Qin, Y., Su, R., Lin, J., Yu, Y., Zhang, W.: Learning
  {ID}-free item representation with token crossing for multimodal
  recommendation (2024)

\bibitem{KZScholarDREAM}
Zhang, K., Qin, Y., Su, R., Liu, Y., Jin, J., Zhang, W., Yu, Y.: {DREAM}: A
  dual representation learning framework for multimodal recommendation (2024)

\bibitem{KZScholarDualAligned}
Zhang, K., Qin, Y., Su, R., Liu, Y., Zhang, W., Yu, Y.: A dual-aligned model
  for multimodal recommendation. In: China Conference on Information Retrieval.
  pp. 14--27 (2024)

\bibitem{DDRM}
Zhao, J., Wenjie, W., Xu, Y., Sun, T., Feng, F., Chua, T.S.: Denoising
  diffusion recommender model. In: Proceedings of the 47th International ACM
  SIGIR Conference on Research and Development in Information Retrieval. p.
  1370–1379. SIGIR '24 (2024)

\bibitem{KZScholarProcessRewardSurvey}
Zheng, C., Zhu, J., Ou, Z., Chen, Y., Zhang, K., Shan, R., Zheng, Z., Yang, M.,
  Lin, J., Yu, Y., Zhang, W.: A survey of process reward models: From outcome
  signals to process supervisions for large language models (2025)

\bibitem{CCFCRec}
Zhou, Z., Zhang, L., Yang, N.: Contrastive collaborative filtering for
  cold-start item recommendation. In: Proceedings of the ACM Web Conference
  2023. p. 928–937. WWW '23 (2023)

\bibitem{MWUF}
Zhu, Y., Xie, R., Zhuang, F., Ge, K., Sun, Y., Zhang, X., Lin, L., Cao, J.:
  Learning to warm up cold item embeddings for cold-start recommendation with
  meta scaling and shifting networks. In: Proceedings of the 44th International
  ACM SIGIR Conference on Research and Development in Information Retrieval.
  pp. 1167--1176 (2021)

\bibitem{Heater}
Zhu, Z., Sefati, S., Saadatpanah, P., Caverlee, J.: Recommendation for new
  users and new items via randomized training and mixture-of-experts
  transformation. In: Proceedings of the 43rd International ACM SIGIR
  Conference on Research and Development in Information Retrieval. p.
  1121–1130. SIGIR '20 (2020)

\end{thebibliography}

\end{document}